\documentclass[12pt,preprint2]{aastex}
\usepackage{graphics,natbib}

\shorttitle{Galactic Halo velocity-resolved \ion N5 and \ion O6}

\begin{document}
\newcommand{\ts}{\textsuperscript}
\newcommand{\dg}{$^\circ$}
\newcommand{\up}[1]{$\times$10\ts{#1}}
\renewcommand{\O}{\ion{O}{6}}
\newcommand{\N}{\ion{N}{5}}
\newcommand{\C}{\ion{C}{4}}
\newlength{\cw}

\title{\ion O6, \ion N5, and \ion C4 in the Galactic Halo: \\ 
     I. Velocity-Dependent Ionization Models }
\author{R\'{e}my
Indebetouw\altaffilmark{1} and J. Michael Shull\altaffilmark{2}}

\affil{CASA, Dept. of Astrophysical and Planetary Sciences, University
of Colorado, 389 UCB, Boulder, Colorado 80309-0389 }

\altaffiltext{1}{present address: Astronomy Department, University of
Wisconsin, 475 N. Charter St, Madison, WI 53706 (remy@astro.wisc.edu)} 

\altaffiltext{2}{also at JILA, University of Colorado and National 
Institute of Standards and Technology (mshull@casa.colorado.edu)}

\begin{abstract}

We explore theoretical models of the ionization ratios of the Li-like
absorbers \ion N5, \ion O6, and \ion C4, in the Galactic halo.  These ions 
are believed to form in nonequilibrium processes such as shocks, 
evaporative interfaces, or rapidly cooling gas, all of which trace the 
dynamics of the interstellar medium.  As a useful new diagnostic, we 
focus on velocity-resolved signatures of several common physical structures:  
(1) a cooling Galactic fountain flow that rises, cools, and 
    recombines as it returns to the disk; 
(2) shocks moving toward the observer; 
(3) a conductive interface with the observer located in the hotter gas.  
This last geometry occurs with the solar system inside a hot bubble, or 
when one looks out through the fragmenting top shell of our local bubble 
blown into the halo as part of the Galactic fountain.  In Paper~II, 
these models are compared to ionization-ratio data from FUSE and 
{\it Hubble Space Telescope}.

\end{abstract}



\section{Introduction}

The nature and dynamics of the interstellar medium (ISM) of galaxies
determines how the energy and matter released by stars are
redistributed through the universe.  It is thus critical to understand
the ISM, and in particular that of the Milky Way, which can be
observed with greater sensitivity and resolution than other galaxies.
The ISM in the disk of our galaxy consists of several phases, from
dense and cold molecular gas to hot and fully ionized.  Stars heat and
disperse the dense clouds in which they form, and that hot gas cools
and recombines eventually to complete the cycle and form more stars.
It is particularly interesting to consider the interface between this
multiphase medium and intergalactic space, where hot gas is released
from supernovae to several kiloparsecs in altitude, forming a hot
diffuse medium first proposed by \cite{spitzer56} as the Galactic corona.

The Galactic corona or halo is almost certainly a dynamic object.  It
is difficult to construct static halo models, because models supported
by thermal pressure are thermally unstable.  If conduction is
sufficiently important to stabilize small-scale instabilities, then
the entire (nearly isothermal) halo is unstable to collapse or
expulsion as a wind \citep[see][for stability
arguments]{bregman80,field65}.  Cosmic-ray supported static halos have
been proposed \citep[e.g.,][]{boulares90}, but there are considerable
uncertainties as to how cosmic rays are confined by the Galaxy (the
Galactic magnetic field topology in particular), and these models do
not explain the high and intermediate velocity clouds.  Considering
these things, \citet{shapfield76} first proposed a ``Galactic
fountain'' of supernova-heated gas rising buoyantly above the disk
until it cools and falls back to the disk.  Smooth Galactic fountain
models have been constructed by several authors. In particular,
\citet{bregman80} discussed the issues of radial flow in the Galactic
gravitational field for a supersonic hot flow which reaches several
kiloparsecs height. More recent models \citep{houck90,breit93} consist
of transsonic flows, which require cooler initial temperatures and
only rise to 1--2~kpc. These models can help to explain the population
of intermediate velocity halo clouds, believed to exist at those lower
heights and velocities compared to the more distant high velocity
clouds \citep[e.g.,][]{hivcdist}.  The cooling layer certainly
fragments into small clouds by Rayleigh-Taylor instability even if no
other inhomogeneities exist \citep[e.g.,][]{berry98}.

Superbubbles, worms, and shell-like structures have been observed in
\citep[e.g.,][]{heiles84}
the upper disk/low halo (few hundred parsecs altitude), but it is
difficult to ascertain whether the hot gas is being expelled into the
halo.  A related issue is the filling factor of hot gas in the disk,
which determines the rate at which hot gas can escape into the halo.
Estimates for that filling factor and for the rate of reheating by
halo supernovae are based on the evolution of supernova
remnants in smooth media, but our current understanding of the ISM is
increasingly inhomogeneous and dynamic, so those arguments are
probably of limited value \citep{kahn98}.  Recently, the combination
of good observations, sophisticated numerical models, and sufficient
comprehension of the Galactic fountain may be beginning to allow
identification of fountain-like rising structures \citep{avillez01}.

The dynamics of interstellar gas in general and specifically in
the Galactic halo may be best understood by studying the hot phase of
the ISM (coronal gas at millions of degrees), and gas at temperatures
$\sim$10\ts{5}~K intermediate to the hot phase and cooler phases
(10\ts{4}~K).  Of the different interstellar gas phases, the coronal
gas is most directly linked to the main sources of energy in the ISM,
supernovae and stellar winds.  Slightly cooler gas is most closely
linked to transient and dynamical processes.  This gas is typically
short-lived because the cooling time is short at at 10\ts{5}~K.
The lithium-like ions of common metals, in particular \ion O6, \ion N5, 
and \ion C4, are sensitive tracers of interstellar gas at several times 
10\ts{5}~K.  The resonance absorption lines of these ions 
are observable with ultraviolet spectrographs on the 
{\it Far Ultraviolet Spectroscopic Explorer} (FUSE), 
and the {\it Space Telescope Imaging Spectrograph} (STIS) and 
{\it Goddard High Resolution Spectrograph} (GHRS) aboard the
{\it Hubble Space Telescope} (HST).  

Section \ref{hotmodels} describes previous models of high-ion column 
densities.  Section \ref{interpretation} describes new models of the 
dynamical signatures of the Li-like ions and their interpretation.
We summarize in \S~4.

\section{Previous models of Li-like ions in the dynamic ISM}
\label{hotmodels}

The ions \ion O6, \ion N5, and \ion C4 are predicted in models of
transient phenomena such as shocks or interfaces between different
temperature gas, with conductive heating, turbulent mixing, or rapid
radiative cooling \citep[see][for a summary]{spitzer96}.  Notable,
especially for \ion O6, is the importance of collisional ionization;
by contrast, \ion{Si}{4} is produced in such transient phonomena, but
is also commonly produced by photoionization in the Galaxy.  The
column densities and column density ratios of these high ions can be
used to determine which physical scenario is predominant in the ISM.
The distributions of \C\ and \N\ in the Galaxy are
(0.6--3)\up{13}~cm\ts{-2}~kpc\ts{-1} and
(0.5--1)\up{13}~cm\ts{-2}~kpc\ts{-1} respectively \citep{SM87}.  The
distribution of \O\ is 1--2 regions per kpc, each with a column
density of (2--5)\up{13}~cm\ts{-2}.  These statistics from
\citet{sheltoncox} came from reanalysis of the {\it Copernicus}
dataset presented by \citet{jenkinsb}, who found a higher frequency of
lower column-density features.  The range in the column density ratio
log[N(\C)/N(\O)] has been previously measured at -1.5 to -0.5 in the
disk and -0.5 to +0.5 in the halo \citep{spitzer96}.  The values of
log[N(\N)/N(\O)] reported by \citet{3c273} are -0.4 to -0.9 in the
halo, but these represent a small heterogeneous sample of the
different lines of sight analyzed by different authors.

The simplest model of gas containing Li-like ions is moderately hot
gas in collisional ionization equilibrium (CIE); \C, \N, and \O\ peak
in ionization fraction at temperatures of $\sim$1, 2, and 3\up{5}~K,
respectively \citep[e.g.,][]{shullvan,sutherland93}.  The ratio
between any two species implies a CIE temperature.  If derived from
two different ion ratios, this CIE temperature does not agree for most
real observations of the Galactic halo, indicating that the halo is
probably not in collisional ionization equilibrium. This is seen in
Figure 1, in which the CIE line ratios lie far to the lower left of
more realistic, nonequilibrium models.  Although there is some
disagreement as to how well atomic abundances are known
\citep{holweger}, recent changes in the solar abundances of carbon and
oxygen are insufficient to bring these two derived temperatures into
agreement, and in fact the latest C/O \citep{allende02} is very close 
to older values.

\begin{figure}
\plotone{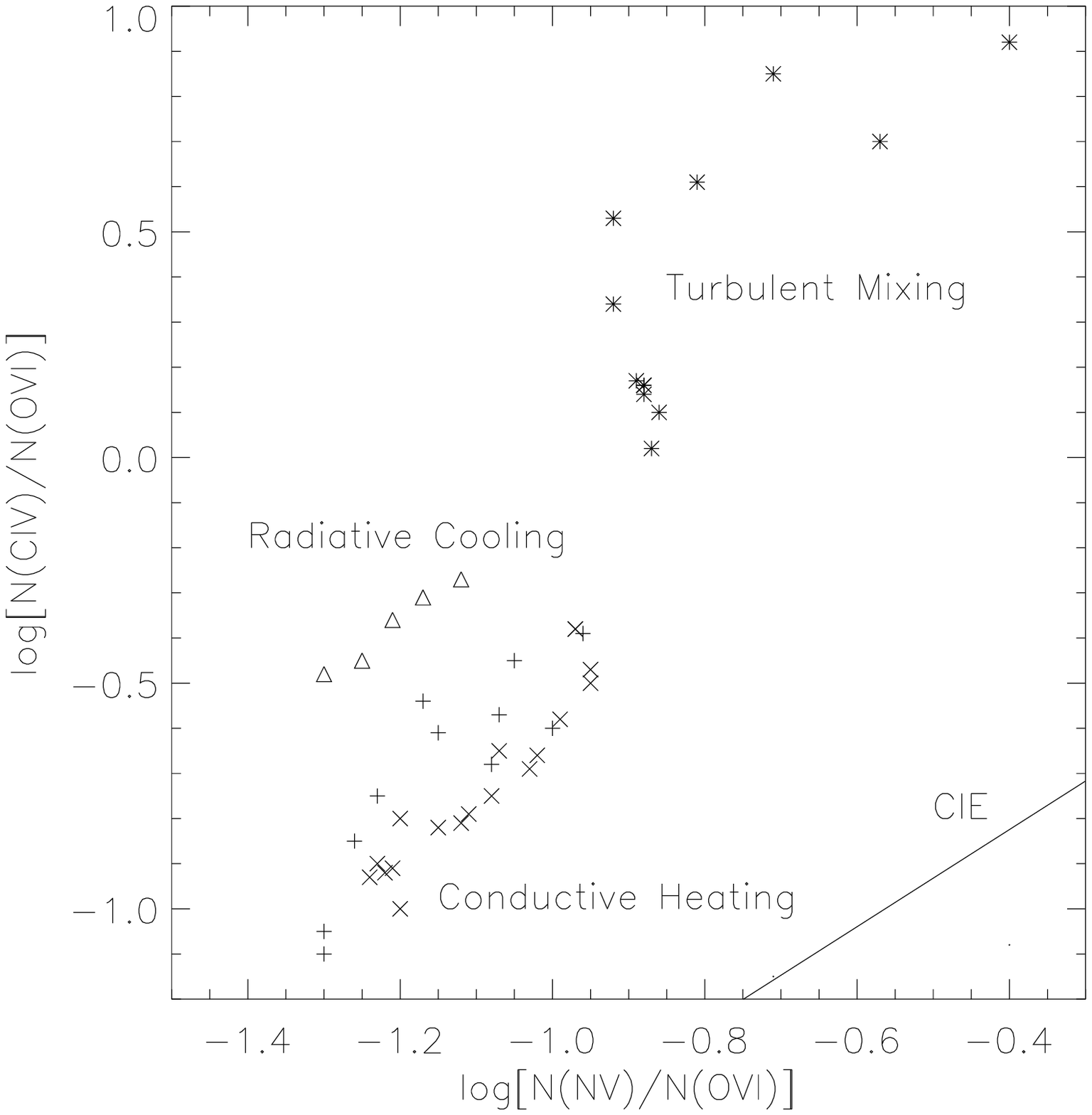}
\caption{\label{diagnostic}
\footnotesize
Line ratios for various models of high-ion production.  The data
plotted are from models in the literature: radiative cooling of
galactic fountain gas \citep[triangles,][]{shapiro93,benjamin93},
turbulent mixing layers \citep[stars,][]{mixing}, and conductive
heating and evaporation of spherical, \citet{bohringer87} and planar,
\citet{borkowski90} clouds (``x''s), and in cooling supernova remnant
shells \citep[pluses,][]{slavincox93,shelton98}.  Also shown are the
ion ratios for hot gas in collisional ionization equilibrium
\citep[CIE, solid line,]{sutherland93}.}
\end{figure}

It is important to understand what the predicted ion column density
ratios are in more dynamic physical models, beginning with conductive
interfaces.  \citet{ballet86} modeled the evaporation of evaporating
spherical clouds with nonequilibrium ionization.  They presented
results for a small ($R$ = 5~pc) isolated cloud in a \citet{mo77} type
model of cool cloudlets in a hot medium of large filling factor.
Relative to calculations of evaporation with collisional ionization
equilibrium, the total column densities of the Li-like ions is
increased, as is the distance from the cloud at which those species
reach their maximum abundance.  Both effects result from delayed
ionization (up to the Li-like ionization stage but especially up to
the He-like stage) as the gas is conductively heated.  Column
densities for all modeled structures are listed in Table
\ref{modeltable}, and column density ratios are given in Table
\ref{modelratios}.  \citet{bohringer87} also modeled evaporating
spherical clouds with nonequilibrium ionization, but self-consistently
included the effects of radiative cooling on the temperature structure.
They found somewhat lower ionization states than \citet{ballet86}, with
higher \N/\O\ and \C/\O.   

\begin{deluxetable}{lccccl}
\setlength{\cw}{3em}
\tabletypesize{\footnotesize}
\tablecaption{\label{modeltable} Li-like ion column densities. }
\tablehead{
\colhead{Physical Situation} &
\colhead{\parbox{\cw}{\begin{center}\ion{Si}{4}\\(10\ts{12}~cm\ts{-2})\end{center}}} &
\colhead{\parbox{\cw}{\begin{center}\C\\(10\ts{12}~cm\ts{-2})\end{center}}} &
\colhead{\parbox{\cw}{\begin{center}\N\\(10\ts{12}~cm\ts{-2})\end{center}}} &
\colhead{\parbox{\cw}{\begin{center}\O\\(10\ts{12}~cm\ts{-2})\end{center}}} &
\colhead{Reference}
}
\startdata
observed sightlines, kpc\ts{-1} & 2--10\tablenotemark{a} & 6--30\tablenotemark{a} & 
5--10\tablenotemark{a} & 10--50\tablenotemark{b} & -- \\
\tableline
evaporating cloudlet    & --         & 1.2--1.5  & 0.5--0.6   & 9--12     & \cite{ballet86}\\
evaporating cloudlet    & .10--.14   & 2.7--3.8  & 1.0--1.2   & 12--14    & \cite{bohringer87}\\
planar conduction front & .10--.16   & 1.6--3.2  & .6--1.0    & 8--10     & \cite{borkowski90}\\
planar conduction front & .029--.097 & .89--2.7  & .40--1.0   & 6.7--14   & \cite{slavin89} \\
stellar wind bubble     & .21--.25   & 3.3--4.0  & 1.3--1.6   & 21--25    & \cite{weaver77} \\
SNR bubble              & .4--.6     & 6.3--10   & 3.2--5.0   & 40--79    & \cite{slavincox92} \\
SNR bubble              & $\sim$0.52 & $\sim$7.8 & $\sim$3.6  & $\sim$47  & \cite{slavincox93}\\
halo SNR bubble         & --         & 8--15     & 3.4--7.9   & 35--150   & \cite{shelton98} \\
\tableline
4 M$_\odot$ cooling     & 3.3--6.4   & 43--79    & 28--36     & 580-600   & \cite{edgarc86} \\
40~pc cooling cloud     & $\sim$25   & $\sim$50  & $\sim$13   & $\sim$200 & \cite{benjamin93} \\
\tableline
turbulent mixing layer  & .0010--.47 & .025--6.8 & .0022--.32 & .017--.81 & \cite{mixing} \\
\tableline
white dwarfs            & 1.4--4.4   & 25--77    & 3.7--12    & 5.6--20   & \cite{dupree83} \\
\enddata
\tablecomments{ Predicted column densities vary with physical
conditions, for example the temperatures of the hot and cold media in
interface models, and an indicative range of column densities is given
for each physical model.}
\tablenotetext{a}{\cite{SM87}.}
\tablenotetext{b}{\cite{sheltoncox,jenkinsb}.}
\end{deluxetable}

\begin{deluxetable}{lccl}
\setlength{\cw}{8em}
\tabletypesize{\footnotesize}
\tablecaption{\label{modelratios} Li-like ion ratios.}
\tablehead{
\colhead{Physical Situation} &
\colhead{$\log\left[{{\rm N(C~IV)}\over{\rm N(O~VI)}}\right]$}&
\colhead{$\log\left[{{\rm N(N~V)}\over{\rm N(O~VI)}}\right]$}&
\colhead{Reference}
}
\startdata
observed halo sightlines & -0.5 to +0.5  & -0.4 t -0.9    &\cite{spitzer96,3c273}\\
\tableline                                                                                 
evaporating cloudlet     & -0.95         & -1.35          &\cite{ballet86}    \\
evaporating cloudlet     & -0.60         & -1.07          &\cite{bohringer87} \\ 
planar conduction front  & -0.4 to -0.8  & -0.9 to -1.1   &\cite{borkowski90} \\ 
planar conduction front  & -0.7 to -0.9  & -1.1 to -1.2   &\cite{slavin89}    \\ 
stellar wind bubble      & -0.8          & -1.2           &\cite{weaver77}    \\ 
SNR bubble               & -0.8 to -0.9  & -1.1 to -1.2   &\cite{slavincox92} \\ 
SNR bubble               & $\sim$-0.8 &  $\sim$-1.1       &\cite{slavincox93} \\ 
halo SNR bubble          & -0.6 to -1.1  & -0.9 to -1.2   &\cite{shelton98}   \\ 
\tableline                                                                      
4 M$_\odot$ cooling      & -0.88 to -1.1 & -1.2 to -1.3   &\cite{edgarc86}    \\ 
40~pc cooling cloud      & -0.4 to -0.5\tablenotemark{a} & -1.2 to -1.4 &\cite{benjamin93}  \\ 
\tableline                                                                                 
turbulent mixing layers  & +0.1 to +0.9  & -0.4 to -0.9   &\cite{mixing}      \\ 
\tableline                                                                      
WD Str\"{o}mgren spheres & +0.6 to +0.7  & -0.18 to -0.22 &\cite{dupree83}    \\ 
\enddata
\tablenotetext{a}{Much lower \C/\O\  ratios ($\sim-$1.7 in log) are present in the
10\ts{6}~K gas before it rapidly cools, but the cooler, overionized stage is
much longer lived.}
\end{deluxetable}

\citet{borkowski90} presented more sophisticated models of a
plane-parallel interface between hot and cold gas.  They modeled the
time-dependent evolution through an evaporative, steady-state, and
recondensation phase. Numbers presented in Tables \ref{modeltable} and
\ref{modelratios} are for the longest, steady-state phase, but much
higher values of N(\N)/N(\O)\ and N(\C)/N(\O)\ are obtained during the
initial evaporation, as the species with lower ionization potentials
ionize up to the Li-like stages faster.  These authors also included a
magnetic field, which can suppress thermal conduction if oriented
parallel to the interface.  The total column densities of all species
are lower in this case, but the N(\C)/N(\O)\ and N(\N)/N(\O)\ ratios
also decrease, because the evaporated gas is hotter, and the overall
ionization state of the front is increased.  This effect was confirmed
quantitatively by \citet{slavin89}.

\citet{weaver77} modeled a conductive interface of a different kind,
at the shell of a wind-blown bubble around a massive star.  The hot
gas on the bubble interior evaporates the cooler shell that
forms around the structure after the swept-up gas is sufficiently
dense to cool.
\citet{slavincox92} performed a similar calculation for a SNR bubble
and found higher total column densities, as expected for that more
energetic phenomenon, but nearly the same ion ratios, as expected for
the same physics.
\citet{slavincox93} explored the same type of models in more detail
and obtained similar results, with scaling laws.
Finally, \citet{shelton98} extended these SNR models and applied them
to a halo with a lower ratio of thermal to nonthermal pressure.  She
found a variation of the column densities and their ratios
with time as the remnant evolves, with a significant decline in the
ionization state after 10\ts{7} years, as the shell slows down and the
interior cools.

\citet{edgarc86} modeled the nonequilibrium ionization of cooling gas
and scaled their results to a Galactic fountain with
4~M$_\odot$~yr\ts{-1} of cooling matter.
More detailed models of cooling gas incorporating radiative
self-ionization \citep{benjamin93,shapiro91,shapiro93} have a similar
range of Li-like ion production.
\citet{mixing} modeled turbulent mixing layers between million degree
gas and cooler (10\ts{2}--10\ts{4}~K) gas.  \O, \N, and \C\ are
produced both in the cooling, overionized hot gas, and in the heating,
underionized cool gas.  These models are physically interesting
because the topology of the interstellar medium includes many
interfaces between hot and cold gas moving at substantial relative
velocities.  The models can produce a large range of ion column
density ratios and generally produce smaller total column densities
per interface than the other models.  However, a number of the model 
parameters are unconstrained by the current state of the observations.

Although \C\ and \ion{Si}{4} can be produced by photoionization near
hot stars and somewhat in the ISM by ambient stellar ionizing radiation
and the soft X-ray background \citep[e.g.,][]{cowie81,black80}, \O\
and \N\ are almost definitely collisionally ionized in the Galactic
disk and halo.  Nevertheless, there are some models that produce these
ions by photoionization near a sufficiently hot source.  \cite{dupree83}
found that hydrogen-rich white dwarfs can produce some \O\ in their
Str\"{o}mgren spheres and in diffuse regions if the circumstellar
medium is patchy.  The Li-like ion column densities depend on impact
parameter of a line of sight with the Str\"{o}mgren sphere, and also
increase rapidly with increasing average ISM density (values in Table
\ref{modeltable} are for $n_H$ = 0.1--1.0~cm\ts{-3}).

Line ratios for model production mechanisms for Li-like ions are
plotted in Figure \ref{diagnostic}.  Clearly, measurements of these
line ratios can help to distinguish between different models.
Additional detail may even be possible -- for example, in the
conductive interface models of \citet{bohringer87} and
\citet{borkowski90} (``x''s in Figure \ref{diagnostic}), models with
increasing magnetic field suppressing the conduction are located
farther to the lower left of the diagram.

It should be noted that the interstellar abundances of the atoms whose
ions are being studied here are not well determined.  The models
presented above are for solar abundances, reported in \citet{allen00},
mostly from the work of \citet{grevesse96}, measured in meteorites and
in the solar atmosphere.  This technique is particularly difficult for
C, N, and O.  The models given later in this paper build on the models
above, and so the implicit metallicities are those of
\citet{grevesse96}.  There is the additional uncertainty of whether
solar abundances are appropriate for the general interstellar medium:
it has been proposed that the Sun may have uncharacteristically high
elemental abundances \citep[e.g.,][]{mathis96}, and that the lower
mean abundances in B stars may be more appropriate
\citep{cuhna92,cuhna94,holmgren90}.  
The recent downward revision of
the solar C, N, and O abundances may bring the interstellar standard back
closer to solar \citep{sofia01,allende01,allende02}.  
Gas in the Galactic halo
in particular has less than solar abundances, although if the hot gas
being observed is recently ejected from the disk, it is less clear
what adjustment needs to be made. If elements are depleted onto dust
grains, this will most significantly affect N(\ion{Si}{4}). It will
also somewhat decrease N(\C), and N(\C)/N(\O)\ \citep[see discussion
in][in the specific context of conductive interface
models]{slavincox92}.

\section{Velocity-Resolved Models of Li-like Ions}
\label{interpretation}

The Li-like ions \C, \N, and \O\ are produced in nonequilibrium
physical situations, such as shocks, conductive interfaces, and rapidly
cooling gas (\S~\ref{hotmodels}).  The different ions are generally
produced at different flow velocities in the structure, so a viewer
looking through such a structure will see trends in the ion column
density ratios as a function of line-of-sight velocity.  The
ratio-velocity signature in relevant physical situations
must be considered to properly interpret trends in the data.
Such models are presented in the following pages.
For conductive interfaces, we took existing models and calculated the velocity 
signature that would be observed along a sightline passing through the interface.
For radiative shocks, we ran the shock code of J. Raymond and calculated the 
velocity signature.  For supernova remnants, we consider the velocity-resolved 
results previously presented by \citet{shelton98}.  Finally, for a Galactic fountain, 
we integrated the steady-state equations published by \citet{houck90}, 
coupled with the recombination code of \citet{bbc}, to determine the ionization structure 
as a function of line-of-sight velocity.

\subsection{Conductive Interface}

\begin{figure}
\plotone{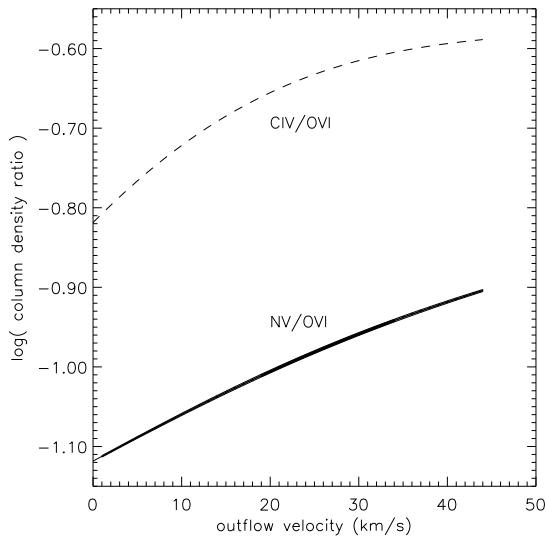}
\caption{\label{bohringer-vel}
\footnotesize
 Column-density ratio as a function of outflow velocity calculated
 from the model of \cite{bohringer87} of an evaporating cold cloud in
 a hot medium.  An observer in the cool cloud would see this ion ratio
 spectrum, with positive velocities moving away from the observer.
 The thickness of the \N/\O\ line scales with the density at that
 point in the flow. Of the sightlines in this study with all three ions,
 some have \N/\O\ profiles that can be explained by a conductive
 interface, but the models predict lower \C/\O\ than observed.}
\end{figure}

One of the oldest
interface models containing the Li-like ions is that of a conductive
interface.  Figure \ref{bohringer-vel} shows the log of column density
ratios as extracted from the models of \cite{bohringer87}. 
We created this figure by extracting the Li-like ion fraction, temperature, outflow
velocity, and pressure as a function of radius from their plots and
calculating the column density of the Li-like ions at each radius.
Then, we integrated the column density over radius, with a parcel of
gas at a certain radius modeled in the absorption spectrum as a
Gaussian velocity component centered at the outflow velocity at that
radius and having the thermal width of the gas at that radius.

\cite{borkowski90} present more sophisticated conductive interface
models, with time evolution and a magnetic field.  Although the
authors do not provide spatially resolved ionization information, they
note that the thickness of an ionized layer increases with ionization
stage, with \C\ occupying $\sim$0.3~pc, \N\ 1--3~pc, and \O\
$\sim$8~pc.  During the initial evaporative phase, the velocity
structure shown in their Figure 4 appears similar to the models of
\citet{bohringer87}, and would thus also show a decrease of ionization
stage with increasing outward velocity.  It is unfortunately not
possible to verify this from the limited information presented in
\citet{bohringer87}.  During the steady-state and condensing phases,
there is flow back from the hot medium, and thus for the
longest-lasting phase of the model, increasing ionization towards more
negative velocities would apparently still be observed.  From the
point of view of an observer in the cool cloud, N(\C)/N(\O)\ and
N(\N)/N(\O)\ would increase towards more positive velocities as in the
\citet{bohringer87} models, but the absorption would be seen at
overall negative velocities.  Naturally the situation is reversed for
an observer in the hot medium.

An interesting note is that the models of \citet{borkowski90} predict
increasing linewidths with ionization potential.  Linewidths are
primarily thermal, and the mean temperature at which a certain ion is
found increases with ionization potential.  The effect is much more
pronouced ($\sim$25\% increase in linewidth) from \C\ to \N\ than from
\N\ to \O\ ($\sim$4\% increase).  The observations of \citet{SSL97}
and \citet{SS92} note an increase in line width between \C\ and \N\
along sightlines that pass through the disk and halo of the Galaxy.
New observations (paper II) also show a hint of downturn in the
N(\C)/N(\O)\ ratios in the line wings, consistent with the trend seem
by other investigators for narrower \C\ linewidths than other Li-like
ions.  No such trend is seen in the N(\N)/N(\O)\ ratio, but in
conductive interfaces similar to the models of \citet{borkowski90} and
in gas in collisional ionization equilibrium, the expected linewidth
difference is much smaller between \N\ and \O\ than between \C\ and
\O.

\subsection{Radiative Shock}

Shock waves are ubiquitous in the interstellar medium.  Of particular
interest here are the possible shock at the edge of the local bubble
and shocks in a Galactic fountain flow.  The latter could arise on the
way up, either in the tops of supershells that are bursting out of the
disk, oblique shocks in more complicated structures which are breaking
out, or on the way down, with condensed clouds reshocking against the
diffuse halo as they return to the disk.  There is evidence of \O\ in
high velocity clouds (HVCs), which could result from shocking as they fall
through the halo.  However, if the halo is close to the virial temperature,
then the Mach number of the HVCs is only $M_s\sim$1.2--1.5, marginally
high enough to shock ionize up to \O.

\begin{figure}
\plotone{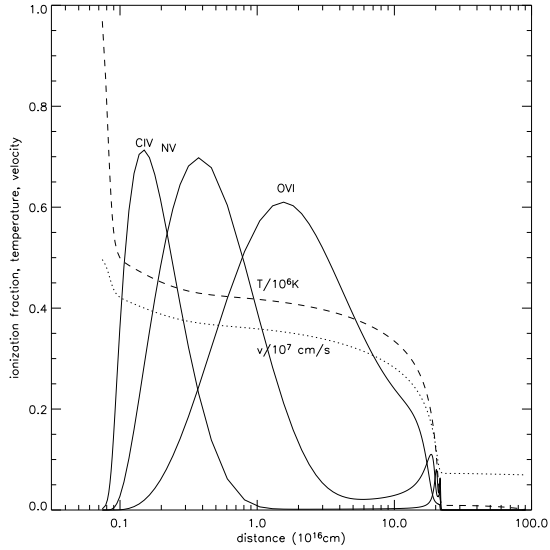}
\plotone{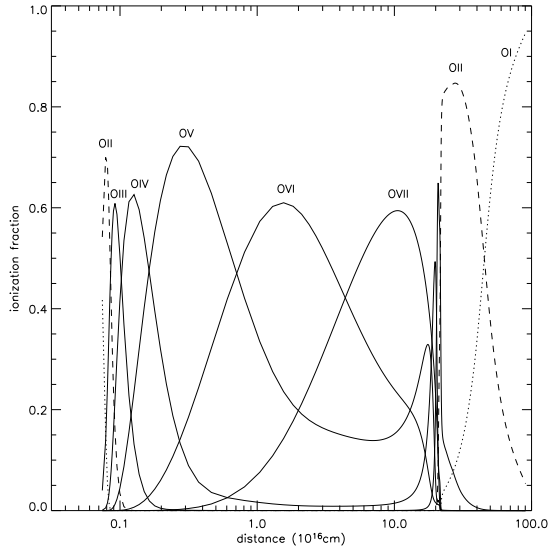}
\caption{\label{shock}
\footnotesize
 Structure of a 200 km~s\ts{-1} interstellar shock, calculated using
 the Raymond shock code.  The first plot shows the Li-like ionization
 fractions of C, N, and O, the temperature, and the flow velocity as a function of
 distance behind the shock.  The second plot shows the ionization
 fractions of all stages for oxygen.  Intermediate and high ionization
 stages (e.g., \ion O5 here) are present immediately behind the shock
 as the gas ionizes up, and farther back in the recombination zone, as
 the gas cools back through $\sim$10\ts{5}~K.  }
\end{figure}

\begin{figure}
\plotone{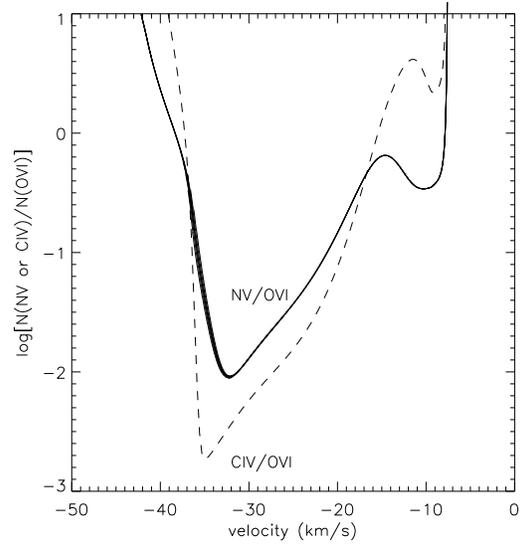}
\plotone{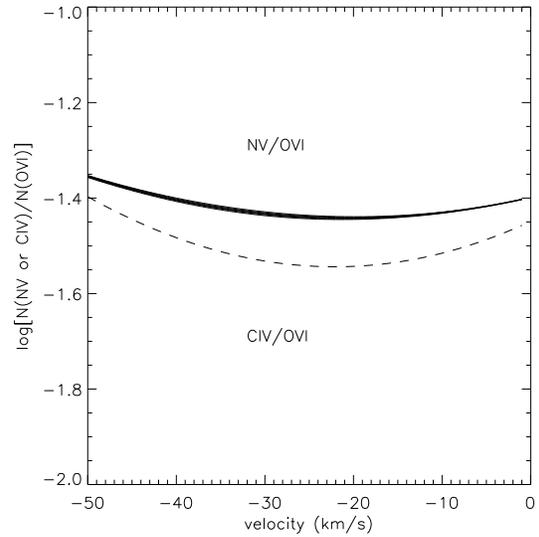}
\caption{\label{shock_no}
\footnotesize
 Ion column density ratios as a function of flow velocity in a 200
 km~s\ts{-1} radiative shock.  The thickness of the line scales with
 the density of \N\ and \O\ present along the line of sight.  The
 first plot shows the column density ratio which would be measured by
 an observer capable of measuring the volume density {\it in situ}.  The
 second plot shows what would be measured by an observer using absorption
 spectroscopy through the shock. The gas at each point in the flow is
 convolved with the thermal width at that point in the flow, greatly
 smoothing the observed structure.  }
\end{figure}

Figure \ref{shock} shows the structure of a 200~km~s\ts{-1} radiative shock,
which we calculated using the Raymond shock code \citep[J. Raymond, private
communication 1998,][]{hrh87}.  Gas immediately behind the shock is
highly underionized compared to its high temperature.  Cooling is slow
above a few million degrees and the gas ionizes up.  Further behind
the shock rapid cooling takes place and the gas recombines back
through the various stages.  There is thus a secondary rise in Li-like
ions in the cooling zone, the degree of which depends on how much of
the He-like ions were produced, which depends on the shock velocity
and the metallicity of the gas.  The details of the
ionization-velocity structure seen by an observer looking through a
shock may change depending on these parameters, but the general trend
is the same, because the bulk of the observed Li-like ions are are
seen as the gas is ionizing up, and \C, \N, and \O\ are ionized
sequentially as the flow velocity slows down.  Figure \ref{shock_no}
illustrates the ion column density ratio as a function of rest-frame
flow velocity, i.e. the line-of-sight velocity seen looking through a
shock moving towards the observer.  As before, the thickness of the
line scales with the density of \N\ and \O\ present along the line of
sight, so although there are parts of the shock in the cooling zone in
which the column density ratio N(\N)/N(\O) increases with velocity, in
the part of the shock with the largest column of those ions, the ratio
is decreasing with velocity.  A negative slope of N(\N)/N(\O) with
increasingly positive velocity would be observed in a sightline
dominated by a a shock moving towards towards the observer, for example
on the front faces of Galactic fountain or high velocity clouds
falling back to the disk and being shocked against the diffuse halo gas.

\subsection{Supernova Remnant Shell}

Another structure which may be particularly relevant to observations
of Li-like ions in the Galactic halo is a supernova remnant shell.
The tops of superbubble shells may be launched into the halo as
partially coherent clouds, and if we are living inside a superbubble,
there may be remnants of the top of our shell above us, through which
many of the sightlines in this study would pass.

\begin{figure}
\centerline{\resizebox{1.2in}{!}{\includegraphics{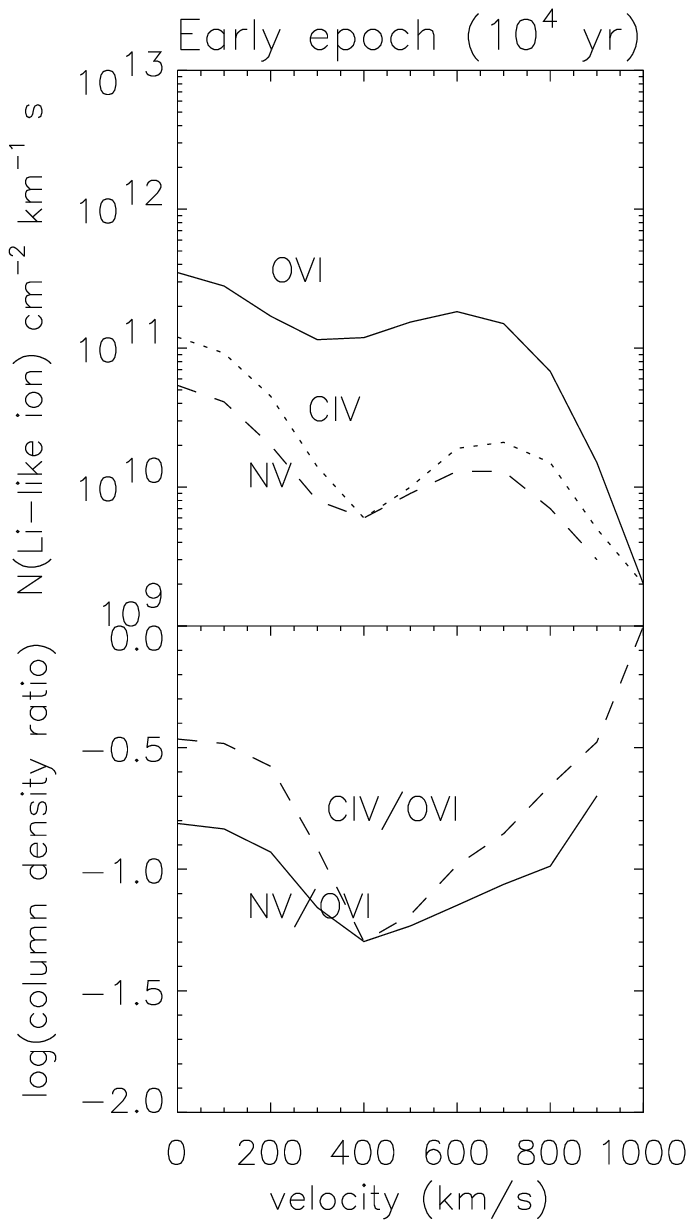}}
\resizebox{1.2in}{!}{\includegraphics{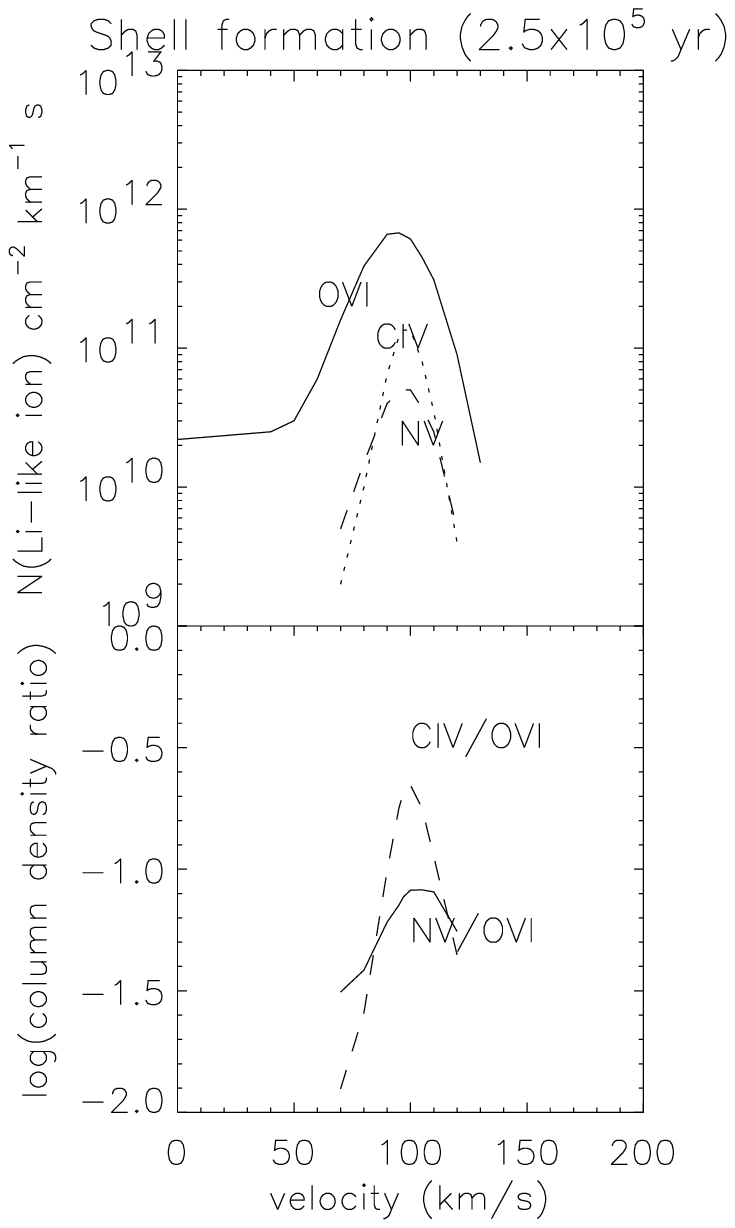}}
\resizebox{1.2in}{!}{\includegraphics{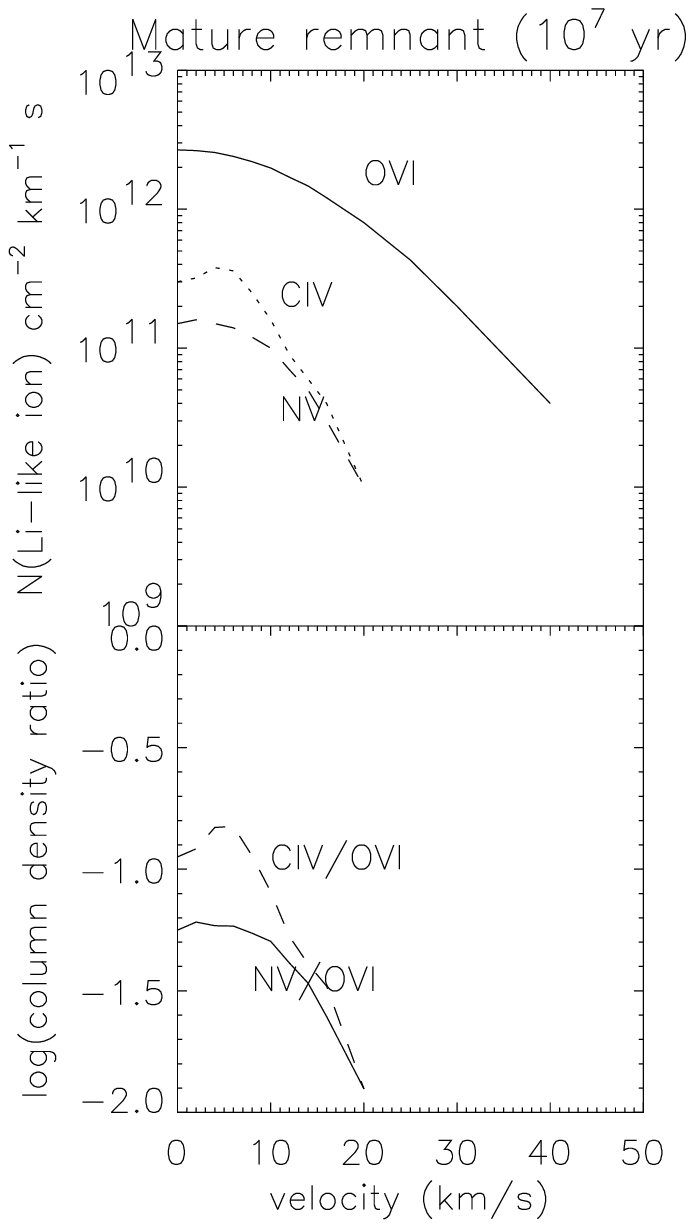}}}
\caption{\label{shelton}
\footnotesize
 Column densities and ratios as a function of velocity for a sightline
 through the center of a supernova remnant, at various stages of the
 bubble evolution (\cite{shelton98}).  The profiles are symmetric about
 zero velocity, and only positive velocities are shown. Note that the
 velocity scales change significantly as the shock decelerates. }
\end{figure}

\cite{shelton98} modeled evolving supernova remnant (SNR) bubbles
expanding in ambient conditions typical of the lower halo.  In
particular, she presented the velocity-resolved column densities of
Li-like ions (her Figure 11).  Those values and the column density
ratios are shown in Figure \ref{shelton}.  Clearly, the ion-ratio
profile depends on the stage of evolution of the SNR.  At very early
stages, the interior of the remnant has not yet ionized past the
Li-like stages, so these species are found both behind the shock and
in the interior, leading to the double-peaked profile in column
density.  Later, the ions are found in the cooling but overionized gas
at the periphery of the remnant.  Finally, as the aged remnant
cools, \O\ and \N\ are found mostly in the interior, and \C\ near the
outskirts.

The profiles of aged remnants are most relevant to the possibility of
our looking out through our own superbubble shell, and these profiles
show a decrease in N(\N)/N(\O) with increasing velocity.
Additionally, if there is a rising shell
above our superbubble, it is probably fragmented owing to
Rayleigh-Taylor instabilities as the superbubble blowout expands through the
density gradient out of the Galactic disk.  Fragmentation and further
disturbance by Kelvin-Helmholtz instability, as the shell passes
through more stationary diffuse gas, would confuse any
velocity-ionization signature.  The nature of our local bubble would
be better tested with spectroscopy towards stars just outside the local bubble,
in which the gas at the bubble interface
could be observed independently from other gas which may be involved in
different physical processes.  The theoretical ion-ratio profiles of
remnants at different ages could be very useful for such local bubble
observations, or other observations in the disk of the Galaxy which
pass through known remnants.  If the geometry is well-enough
understood, observations of the column density ratio could determine
the age of the remnant.

\subsection{Galactic Fountain}

The nature of a Galactic fountain flow is determined by how the
quickly gas cools and recombines as it flows upwards out of the disk.
Quantitatively, the comparison is between the cooling time $\tau_c$
and the sound crossing time $\tau_s$, which is roughly the time to
establish hydrostatic equilibrium:
\begin{eqnarray*}
\tau_s &=& {1\over g}\left({{kT}\over{\gamma\mu m_p}}\right)^{3/2}\\ 
\mbox{\rm and}&& \\ 
\tau_c &=& {{nkT}\over{(\gamma-1)n_en_H\Lambda(T)}},
\end{eqnarray*}
where $\gamma$ is the adiabatic exponent, $\mu$ is the atomic weight,
and $\Lambda(T)$ is the cooling function \citep[e.g.,][who discuss the
nature of the fountain flow for the different values of
$\tau_s/\tau_c$]{houck90}.  If $\tau_c\gg\tau_s$ then a static system
can be established, but it will be unstable to convective overturn and
thermal instability \citep[as noted by][the cooling rate of the static
structure is highest at the top of the halo, so clouds will condense
out from the top down and be formed with small negative
velocities]{bregman80}. If $\tau_c\ll\tau_s$ then a supersonic flow
will occur.  As the gas cools rapidly it is thermally unstable and
will condense into clouds which fall out of the flow.  The scale
height of the gas is then set by the cooling time
$H\simeq\tau_cv\simeq\tau_cc_s$, where $c_s$ is the sound speed at the
base of the fountain, and clouds can be formed with positive (upward)
velocities.  For $\tau_c\simeq\tau_s$, transsonic fountain flows exist
with clouds forming near the adiabatic density scale height.

\begin{figure}
\plotone{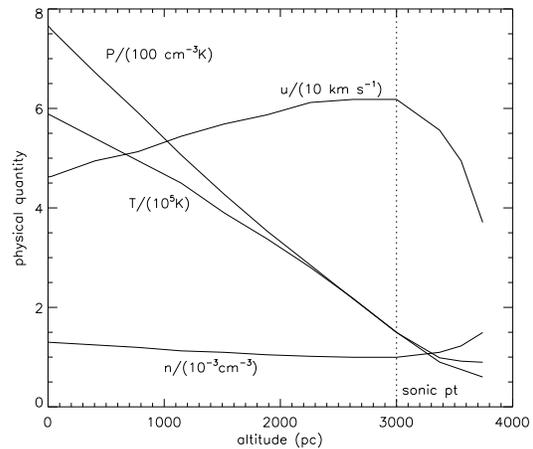}
\caption{\label{hb}
\footnotesize
 \citet{houck90} transsonic fountain.  The vertical dotted line marks
 the sonic point.}
\end{figure}

Modeling of the velocity-ionization structure of a Galactic fountain
will focus here on the transsonic solution ($\tau_s\simeq\tau_c$) of
\citet{houck90}. (As discussed below, flow solutions in which the
cooling and equilibrium times are very different are not as
well-suited to analytic treatment.)  Figure \ref{hb} shows the
structure of the \citet{houck90} fountain flow, an integration of the
one-dimensional steady-state fluid equations, which pass through a
sonic point. The equations have the same topology as a standard
solar/stellar wind \citep{parker58}.  The solution stagnates at the
top, forming an unphysical dense layer, but in this zone of rapid
cooling clouds will condense out with near-zero or small positive
vertical velocities and fall back to the disk.

\begin{figure}
\plotone{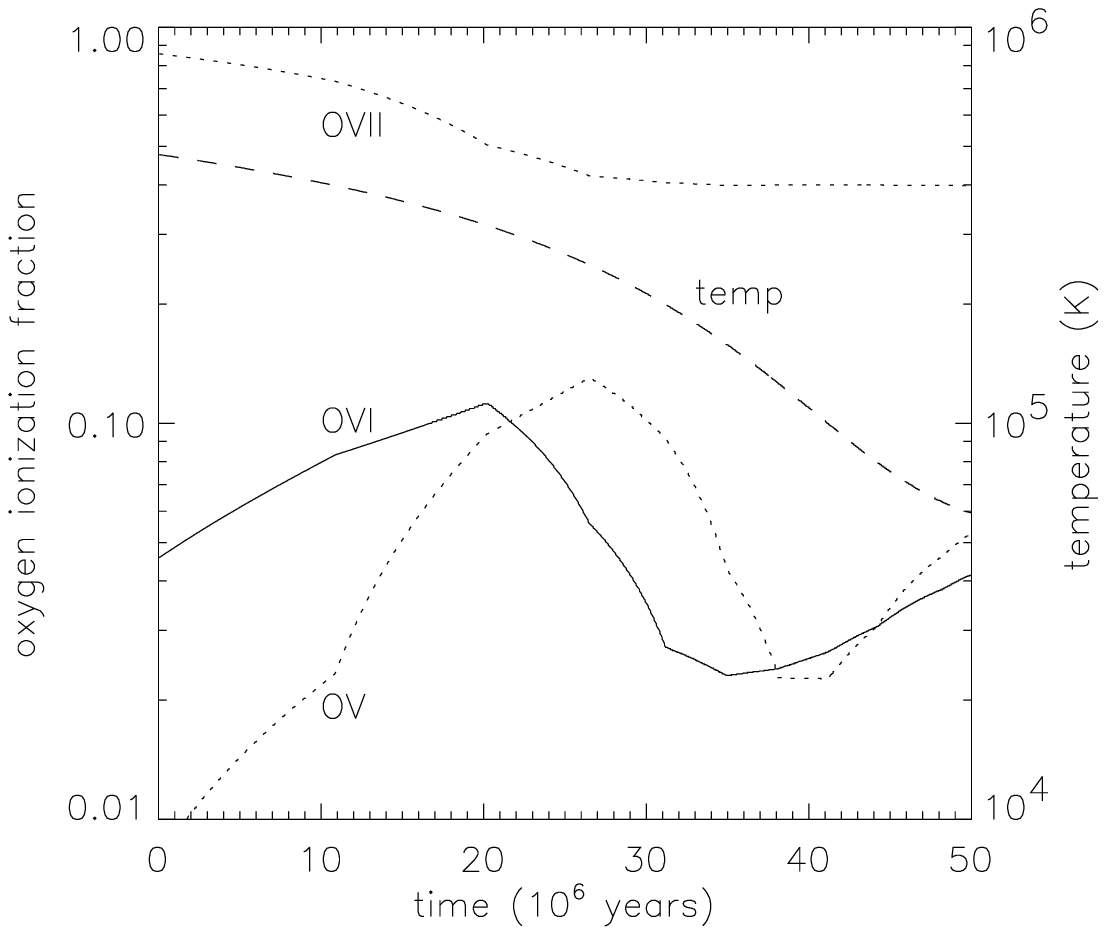}
\caption{\label{dielectric}
\footnotesize
 Oxygen ionization fractions in a nonequilibrium cooling fountain
 flow, calculated using the analytical approximation to the
 hydrodynamic equations of a transsonic fountain of \citet{houck90}
 coupled to the ionization code of \citet{bbc}.  \O\ persists to low
 temperatures because the gas cools much faster than it can recombine.
 The \O\ ionization fraction can even rise at low temperatures due to
 the much-delayed recombination of \ion O7 (see text).  }
\end{figure}

Figure \ref{dielectric} shows the oxygen ionization fractions in a
nonequilibrium cooling fountain flow, which we calculated using the
nonequilibrium recombination code of \citet{bbc}.  We integrated the
recombination rate through the temperature and
electron density given as a function of time and vertical velocity by
the \citet{houck90} transsonic fountain.  It would be slightly more
consistent to integrate the fluid equations with \citet{bbc}
nonequilibrium cooling instead of the \citet{edgar86} nonequilibrium
cooling used by \citet{houck90}, but the differences in the cooling
curves are small enough for the combination to be accurate
(R. Benjamin, private communication 2001).  \ion O7 cannot recombine
fast enough as the flow cools rapidly through its collisional
ionization equilibrium temperature,$\sim$(5--10)\up{5}~K.
Significant \ion O7 remains to low temperatures 
($\lesssim$10\ts{5}~K if the density is low enough), until dielectronic
recombination allows a recombination cascade.  The result is that not
only does \O\ persist to low temperatures because it is ``frozen in''
(it cannot recombine fast enough to keep up with cooling), but there
can be a {\it rise} in the \O\ fraction as the He-like \ion O7
recombines. This happens for the other Li-like ions \N\ and \C\ for
the same reasons.  The magnitude of this effect is dependent on
dielectronic recombination rates which have undergone continual revision
with time \citep{bbc,shapiro76,kafatos73}, and makes the modeling of
Li-like ions in cooled (T$\sim$10\ts{4}~K) clouds uncertain (see
discussion and a crude model below).

\begin{figure}
\epsscale{0.8}
\plotone{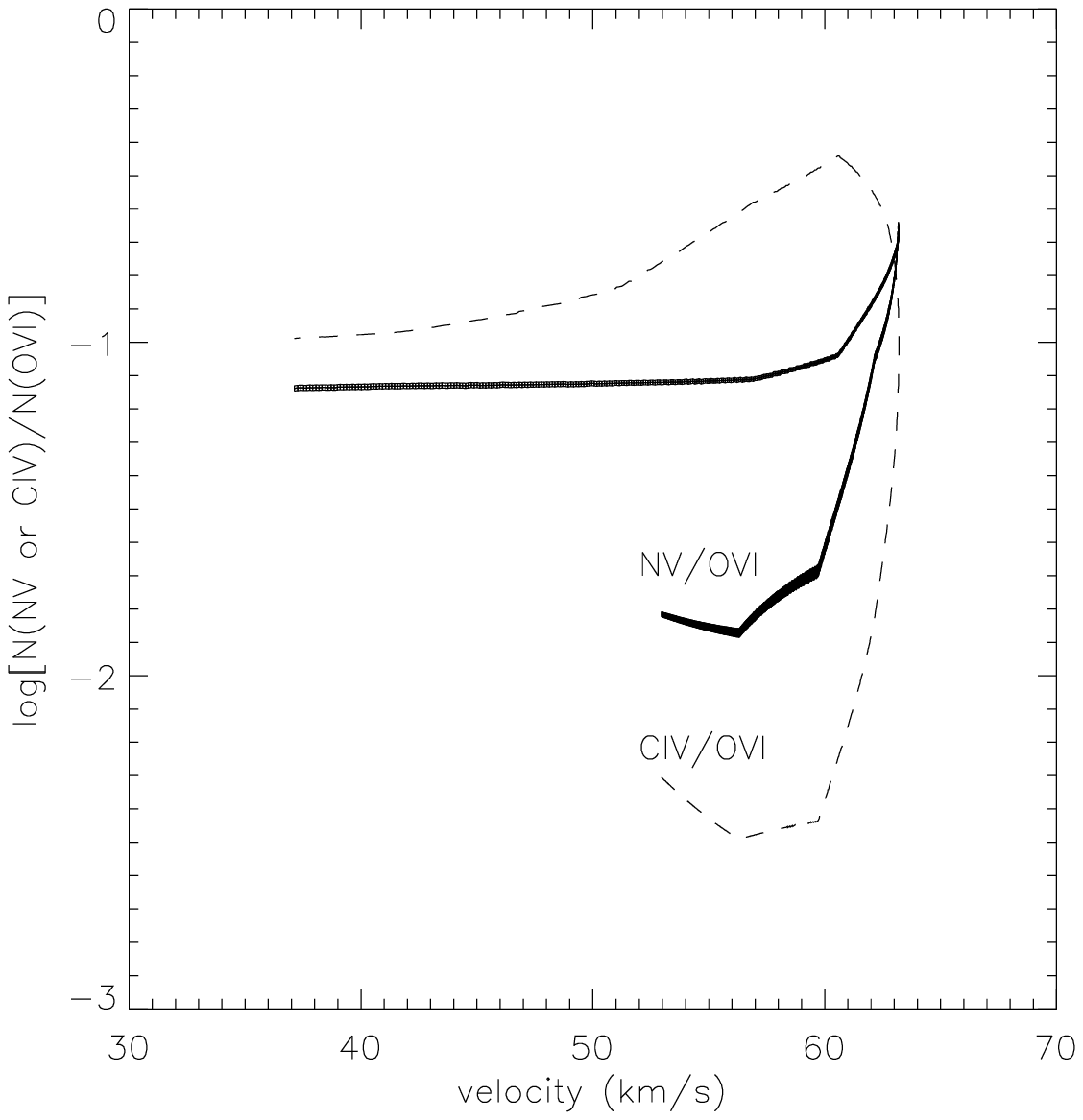}
\plotone{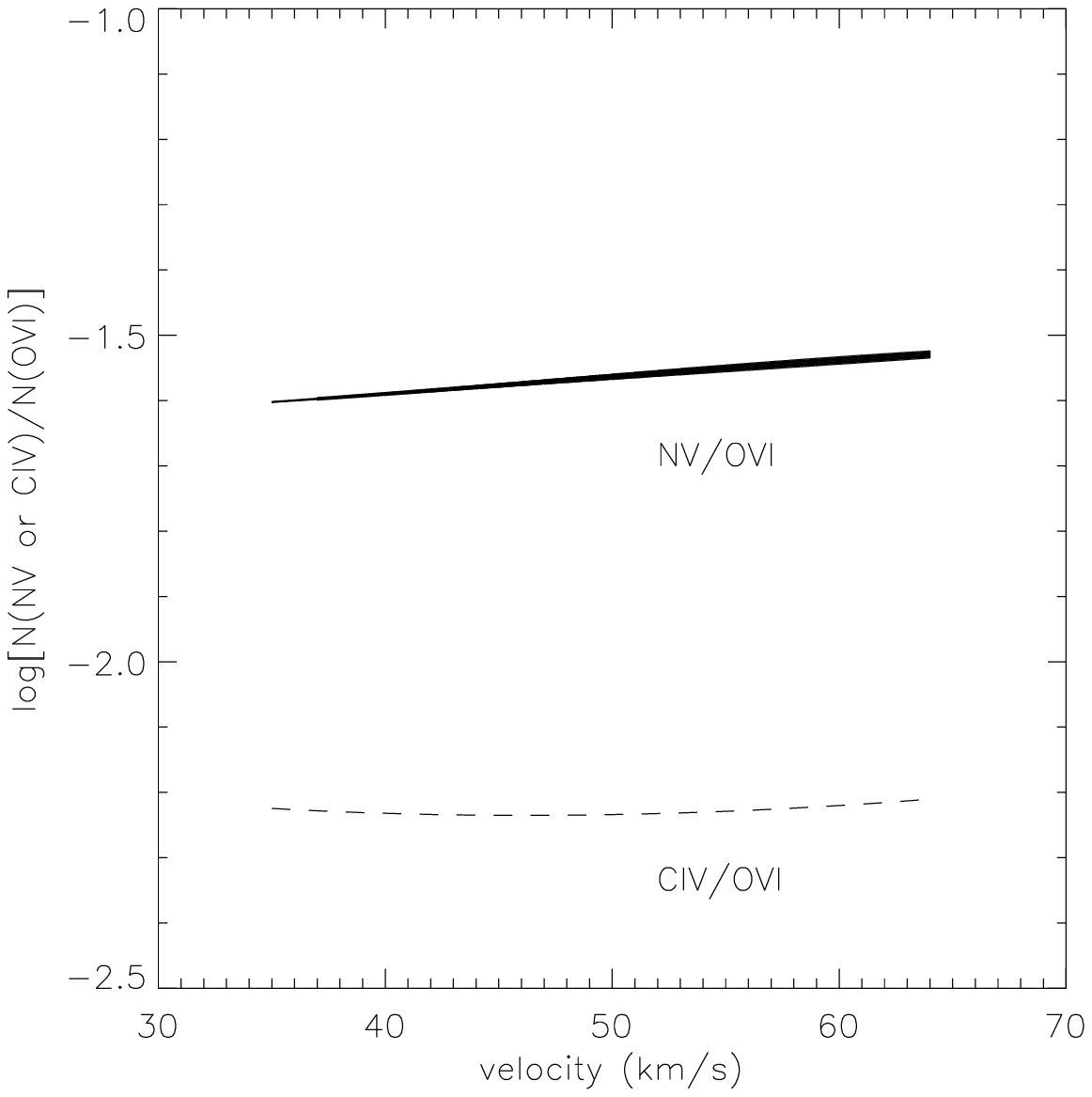}
\epsscale{1.0}
\caption{\label{hbratio}
\footnotesize
 Ion column density ratios as a function of velocity for a
 \citet{houck90} transsonic fountain coupled to \citet{bbc}
 nonequilibrium reionization.  The thickness of the line scales with
 the \N\ and \O\ density along the line of sight, so the observed
 absorption profile is weighted towards the ratio with the thickest
 line.  For clarity, the column density weighting is only shown for
 log[N(\N)/N(\O)].  The first plot shows the ion column density along
 the flow as it would be measured by an observer in situ capable of
 measuring the ion volume density at a point in the flow.  The second
 plot shows what the observed ion column density ratio would be for an
 observed using absorption spectroscopy as was done in this study.
 The ion column density at each point in the flow is convolved with
 the thermal width of the gas at the temperature at that point in the
 flow to create an absorption profile, and the trend is thus
 significantly smoothed out. }
\end{figure}

Figure \ref{hbratio} shows the ion column density ratio calculated by
coupling the \citet{bbc} reionization with the \citet{houck90}
transsonic fountain.  The parameters for this particular model are set 
by choosing the fountain base temperature to be 6\up{5}~K. 
The two branches in the first plot are a result
of the velocity of the fountain accelerating, then decelerating again
after passing through the sonic point.  The horizontal upper branch is
from the rapidly cooled gas at a few 10\ts{4}~K, which will condense
into clouds and fall out of the flow.  The column density of Li-like
ions at this point in the flow is highly uncertain, 
and the column plotted here is probably an overestimate.
(R. Benjamin, private communication 2001).  The width of the line in
the plot scales with the column density of \N\ and \O\ at that point
in the flow.  Thus, an observation through the fountain is dominated
by the rising, cooling gas.  The second plot shows the ion column
density ratio which would be observed using absorption spectroscopy. 
Each parcel of gas in the flow was convolved with the thermal width at
that point in the flow to create an observed absorption profile, and
the ion ratio is calculated from that observed profile.  Since the
transsonic fountain accelerates, N(\N)/N(\O) increases with
increasingly positive velocity. 
However, the trend is very weak due to the small velocity
gradient in this fountain model, and it is not clear that the
transsonic smooth fountain is a realistic description of the patchy
Galactic halo.

It has been noted \citep{avillez00} that there have been two types of
Galactic fountain models in the literature, those with somewhat
smoothly distributed rising gas over the disk, as used here, and
models that focus on the evolution of superbubbles as the fountain
drivers, irregularly spaced in time and galactic location
\citep[e.g.,][]{normanik89}.  Although the latter model is driven
differently and rises to larger Galactic altitude than that of
\citet{houck90}, the cooling and equilibrium times are still
close, the expected spread of velocity in the rising gas is modest,
and so the observed ion column density ratio as a function of velocity
is probably not qualitatively different from the model considered
here.  What is needed is an improvement over either type of simple
model, i.e. full multidimensional time-dependent models of the
fountain with nonequilibrium cooling and nonequilibrium reionization.

Galactic fountain flows in which the cooling and equilibrium times are
very different are less well-suited to an analytic treatment because,
as discussed above, it is difficult to accurately follow the
recombination state of gas which is highly overionized but which has
cooled to $\sim$10\ts{4}~K and thermally condensed.  In a supersonic
fountain ($\tau_c<\tau_s$) clouds condense quickly and observations
are dominated by this uncertain cool phase.  In a highly subsonic
fountain ($\tau_c\gg\tau_s$) there is no velocity signature because
the gas rises and establishes a quasistatic configuration before
cooling through the Li-like ions.  Any difference in ionization as a
function of velocity again occurs in the cold clouds.

\begin{figure}
\epsscale{0.8}
\plotone{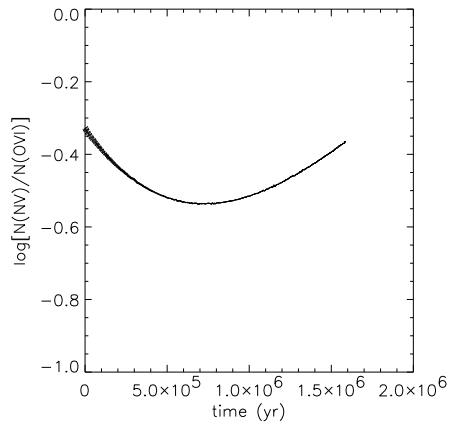}
\epsscale{1.0}
\caption{\label{recomb}
\footnotesize
 Ion column density ratio log[N(\N)/N(\O)] as a function of time in
 recombining cool gas.  Initial ionization fractions are indicative of
 what might be found in gas that has rapidly cooled to a few
 10\ts{4}~K, is highly overionized, and has condensed into a cloud by
 thermal instability.  The exact behavior (for example, the magnitude
 of the rise relative to the drop in ion ratio) depends on the choice
 of uncertain recombination coefficients and even more uncertain
 initial ionization states.}
\end{figure}

Nevertheless, to estimate the behavior of recombination in cool
clouds, Figure~\ref{recomb} shows the ion column density ratio as a
function of time for a crude model.  We consider a cloud
formed by thermal instability in a smooth fountain flow at 10\ts{4}~K
with \N, \ion N6, \O, and \ion O7 ionization fractions of 0.06, 0.01,
0.02, and 0.005 respectively.  These values are fairly uncertain, and
were taken from \citet{shapiro76} with consideration of \citet{bbc}
and \citet{sutherland93} and the known limitations of these models 
(R. Benjamin, private communication 2001).  The cloud cools slowly at
a few $\times$10\ts{4}~K, remaining in pressure equilibrium with the
rising smooth hot halo, and thus the density increases, but not by a
huge factor.  At the same time, hydrogen recombines, so the electron
density must decrease.  Considering the other large uncertainties of
this model, the simple approximation is adequate to allow
hydrogen to recombine down to the hydrogen ionization fractions
$X_H\sim0.1-0.5$ observed in diffuse
halo clouds \citep{spitzfitz93} and modeled in photoionization
equilibrium with ionizing radiation escaping from the Galactic disk
\citep{bland99} or old supernova remnants \citep{slavin00}.  Total
(radiative and dielectronic) recombination rates were mostly taken from
\citep{nahar00,nahar99,nahar97,schippers01}, which
shows significant differences from previous work.

The ion ratio N(\N)/N(\O) first decreases with time, as the \N\ total 
recombination rate is larger than that for \O.  This fact was not true 
for the older rates of \citet{shullvan}.  The column density ratio
then rises as additional \N\ and \O\ recombine down from \ion N6 and
\ion O7 (the recombination rate for the latter is larger).  As
elsewhere, the thickness of the line in Figure \ref{recomb} scales
with the local density of \N\ and \O.  The behavior of the ion ratio
with time (how much it rises at late times, etc.) varies with the
choice of initial ionization fractions, but the recombination time in
the cool cloud is always short compared to the return time for
the ballistically traveling cloud: the cloud velocity in the Galactic
gravitational field ($g\simeq$10\ts{-8}~cm~s\ts{-2}) only changes by 3
km~s\ts{-1} in the 10\ts{6}~year recombination time.  Thus any change
in the ionization ratio as a function of velocity determined by the
distribution of vertical velocities of clouds at their formation time,
not by a slow change in ionization state as the clouds decelerate.
The height and vertical velocity of cloud formation depends on the
initial temperature of the fountain as discussed above, and
simulations show a fairly large spread for a given fountain flow
\citep{bregman80}. Although the general assertion can be made that
clouds recombine as they travel the fountain flow, no conclusions can
be drawn about an observable velocity-ionization signature from this
type of simple considerations of single clouds.  What is needed are
full nonequilibrium 3-d simulations of fountain dynamics.

\subsection{Other Halo Models}

Although the most consistent model of the halo is of a dynamic flow of
gas heated and ionized by (thermal) collisions, it is worth
considering some of the other physics that could be important and how
that might relate to observed Li-like ion column density ratios.

Magnetic fields play a large, but usually neglected, role in
interstellar dynamics.  For example, the magnetic pressure in the
Galactic disk is about the same as the thermal pressure.  Synchrotron
emission from our Galactic halo and the halos of other Galaxies show
that the scale height of the magnetic field is large \citep{han94}.
Dissipation of magnetic energy is a viable mechanism for heating gas
to millions of degrees in the disk \citep{tanuma01,hartquist84},
moderately ionizing and heating diffuse gas in halo \citep{birk98},
and even in producing the X-ray emission of boundary regions of high
velocity clouds passing through the halo \citep{zimmer97}.
\citet{raymond92} presents a quasistatic halo model in which the gas
is heated by magnetic reconnection.  The model can produce reasonable
column densities of high and moderately ionized species, and
reasonable intensities for emission lines of those species.  It may
have some difficulty fitting both the highly ionized species and the
less ionized species, because the lower energy reconnection mechanism
which is invoked requires possibly more magnetic power than is
available in the Galaxy.  In addition, this model would have no
kinematic signature, being quasistatic, and the stability of such a
model is not discussed in \citet{raymond92}.

Cosmic rays are another sometimes neglected but important part of the
interstellar medium. Again, the cosmic ray pressure in the plane is
comparable to the thermal and magnetic pressures.  \citet{boulares90}
presented a simple static halo supported by cosmic rays (CR), and
argued that the stability times are of the order of 10\ts{7}~years.
More recent work calls into question the stability of CR supported
haloes, especially in the presence of CR diffusion which makes CR
supported structures more Rayleigh-Taylor unstable \citep{breit93}.  A
static CR supported halo would also not have variation of ion column
density ratios with line-of-sight velocity. Perhaps more likely is a
CR-driven wind, which may operate in parallel with a Galactic
fountain, driving material out in regions of open magnetic field
topology, while lofted material in regions of closed magnetic field
falls can fall back to the disk \citep{breit91}.  These authors
concentrate on the CR-driven outflow at hundreds of kiloparsecs from
the disk, but their model does have monotonically increasing velocity
as gas rises and cools, entrained in the CR wind.  The gradient of
$\sim$4 km~s\ts{-1}~kpc\ts{-1} is almost identical to that in the
lower part of the \citet{houck90} transsonic fountain, and thus would
have a very similar trend of ion column density ratio with
velocity. There is the possibility that damping of hydromagnetic waves
in the outflow can transfer energy to the gas; the recombination of
the gas could be thus delayed to higher altitudes, where the velocity
gradient is smaller, and perhaps a steeper change in ion column
density ratio as a function of velocity could be produced. To
fully understand the Li-like ions in this CR wind, more specific
discussion of ionization is required than is provided in
\citet{breit91}.

Finally, one can consider that the halo gas is 
partially photoionized.  It has been argued that photoionization is
not negligible in the halo \citep{slavin00,bland99}, but it is
difficult to explain species with large ionization potentials by
photoionization, namely \N\ and \O.  This can be easily shown by
calculating the path lengths required to produce the observed
N(\N)/N(\O) column density ratio and simultaneously the total column,
or the N(\C)/N(\O) ratio.  The conclusion is similar for the
entire data set: ion column density ratios require high ionization
parameters and low space densities, but substantial column densities
then require path lengths of tens or even hundreds of kiloparsecs,
which is unreasonable to fit in the Galactic halo.  In addition, the
two ion ratios N(\C)/N(\O) and N(\N)/N(\O) cannot simultaneously fit a
photoionization model in most cases. The ionization parameter
$U=n_\gamma/n_H$ ($n_\gamma$ is the number density of
hydrogen-ionizing photons) derived from N(\C)/N(\O) is about 0.2 dex
lower than that derived from N(\N)/N(\O), indicative of a different
(collisional) ionization source.

\section{Conclusion and Summary}

We have considered the diagnostic power of velocity-resolved column density
ratios in understanding the Galactic halo.  Column density ratios of
Li-like ions in the Galaxy are useful to diagnose the physical
formation mechanism of the gas and to study the interstellar gas
cycle, and a survey of these ions can reveal general trends. In  
Paper II, we present a survey of sightlines observed with FUSE and HST, 
in which the distribution of N(\N) and N(\O) in the halo does not appear 
to favor a dominant physical production mechanism.  

Here, we have presented models of interfaces and cooling nonequilibrium 
gas, focusing on the velocity-resolved N(\N)/N(\O) signatures. In particular, 
we consider conductive interfaces, radiative shocks and supernova
remnants, and cooling gas in a Galactic fountain.  Typical ion column 
density ratios are summarized in Table~\ref{summary},
along with a typical observed slope observed in FUSE and HST data.
One important type
of hot/cool gas interface for which the ionization-velocity signature
has not been discussed is the turbulent mixing layer \citep{mixing}.
The structure of these Li-like ion producing structures is
uncertain. It is likely that there is a turbulent cascade of eddies
which would wash out any velocity-ionization signature, but there may
possibly be large Kelvin-Helmholtz rolls with some coherent velocity
signature.

\begin{deluxetable}{lr}
\tabletypesize{\footnotesize}
\tablecaption{\label{summary} Summary of models}
\tablecomments{Models predict a slope in N(\N)/O(\O) with line-of-sight
     velocity for observers looking through the structure.  Galactic
     fountain model includes \citet{bbc} cooling }
\tablehead{
\colhead{Model} &
\colhead{Slope of Ratio log[N(\N)/N(\O)]}
}
\startdata
\parbox{3in}{Radiative shock} & $\sim$-0.0015 dex~(km~s\ts{-1})\ts{-1} \\
\parbox{3in}{Mature SNR shell \citep{shelton98}} &
          $\gtrsim$-0.02 dex~(km~s\ts{-1})\ts{-1} \\
\parbox{3in}{\citet{houck90} fountain} & $\sim$+0.002 dex~(km~s\ts{-1})\ts{-1} \\
\tableline
Indicative observed slope & -0.0032+/-0.0022(r)+/-0.0014(s) dex~(km~s\ts{-1})\ts{-1}
\enddata
\end{deluxetable}

The observable velocity-ionization trends are weak, because even very
strong trends are washed out by the large thermal width of the gas at
different parts of the flow.  
These trends could be further complicated by thermal instabilities, which 
likely occur in cooling gas such as in a fountain flow, and can also affect 
shocks and supernova shells.  Fragementation of the gas into parcels with 
complex density and velocity structure, as seen in the time-dependent 
shock model of \citet{sutherland} could complicate the signatures 
described here.

Additional confusion can result when long sightlines pass through multiple
structures.  As we show in Paper II, the dispersion of N(\N)/N(\O), both
integrated and velocity-resolved, indicates that no single
production scenario known to date can completely explain the Galactic
halo.  To truly understand the physical production of Li-like ions in
the halo, one needs to analyze gas in localized areas
of physical space, rather than velocity space.  Absorption
spectroscopy towards many halo stars with close angular separation and
different distances could help to isolate gas at a specific altitude.
Similarly, the gas above known superbubble shells or chimneys could be
isolated.  These observations have a greater chance of distinguishing
between models of hot gas production than observations along long
lines of sight.

\acknowledgments

We thank S. Penton, E. Wilkinson, J. Green, K. Sembach, and B. Savage 
for useful discussions.  R.I. was partially supported during this
investigation by an NSF Graduate Student Fellowship to the University
of Colorado.  J.M.S. acknowledges support from theoretical astrophysics
grants from NASA (NAG5-7262) and NSF (AST02-06042).

\end{document}